\documentstyle[twocolumn,aps,epsf]{revtex}
\begin{document}
\newcommand{\beq}{\begin{equation}}
\newcommand{\eeq}{\end{equation}}

\title{STRUCTURAL SUSCEPTIBILITIES IN TOY MODELS OF
PROTEINS}

\author{Mai Suan Li}

\address{Institute of Physics, Polish Academy of Sciences,
Al. Lotnikow 32/46, 02-668 Warsaw, Poland\\
and Institut f{\H u}r Theoretische Physik, Universit{\H a}t zu K{\H o}ln,
Z{\H u}lpicher Stra{\ss}e 77, D-50937 K{\H o}ln, Germany }

\address{
\centering{
\medskip\em
{}~\\
\begin{minipage}{14cm}
New definitions of the structural susceptibilities based on
the fluctuations of distances to the native state of toy protein
models are proposed.
The calculation of such susceptibilities does not require 
the basin of native state and
the folding temperature can be defined from their peak
if the native conformation is compact.
The number of peaks
in the derivatives of distances to the native state with 
respect to temperature, when plotted versus temperature,
may serve as a criterion for foldability.
The thermodynamics quantities are obtained by Monte Carlo and molecular
dynamics simulations.
{}~\\
{}~
\end{minipage}
}}

\maketitle


Understanding of many aspects of protein folding has been recently advanced
through studies of toy lattice models \cite{Dill,Grosberg}. A more
realistic modelling, however, requires considering off-lattice systems.
In lattice models, the native state is usually non-degenerate and it
coincides with the ground state of the systems. 
In the case of off-lattice models the native state has a zero measure
and delineating boundaries of the native basin in off-lattice systems
is vital for studies
of almost all equilibrium and dynamical properties.
For instance, stability of a protein is determined by estimating the 
equilibrium probability of staying in the native basin: the temperature 
at which  this probability is $\frac{1}{2}$ defines the folding temperature,
$T_f$. 

In most studies, such as in \cite{Irback,Klimov,Veitshans}, 
the size of a basin,
$\delta_c$, is 
declared by adopting a reasonable but ad hoc cutoff bound. In Ref. 
\cite{Onuchic}, for instance, the folding kinetics are studied 
by monitoring the number
of native contacts. The definition of the native contacts remains,
however, ambiguous because it depends on a choice of the cutoff distance.
We have developed two systematic approaches to delineate the native basin
\cite{Li1}.
One of them is based on exploring the saddle points on selected trajectories
emerging from the native state. 
In the second approach, the basin is determined
by monitoring random distortions in the shape of the protein around the
native state.
It should be noted that the implementation of these methods becomes difficult
in the case of long chains. The question we ask in the present paper is
what one can learn about the folding thermodynamics and the
foldability of the off-lattice
sequences without the knowledge of $\delta_c$.

We start our
discussion by introducing the following distances to the native state
\begin{eqnarray}
\delta_{d}  \; &=& \; \sqrt{\frac{2}{N^2-3N+2} \sum_{i \neq j,j\pm 1}
(d_{ij} - d_{ij}^{NAT} )^2} \; , \nonumber\\
\delta_{ba} \; &=& \; \sqrt{\frac{1}{N-2} \sum_{i=1}^{N-2}
(\theta_i - \theta_i^{NAT})^2} \; , \nonumber\\
\delta_{da} \; &=& \; \sqrt{\frac{1}{N-3} \sum_{i=1}^{N-3}
(\phi_i - \phi_i^{NAT})^2} \; .
\end{eqnarray}
Here $d_{ij}=|\vec{r}_i-\vec{r}_j|$ are the monomer to monomer distances in the
given structure, $N$ is a number of beads. The subscripts $d, ba$ and $da$
refer to the distances, the bond angles and the dihedral angles, 
respectively.
The superscript $NAT$ corresponds to the native state.
The bond angle, $\theta_i$, is defined as the angle between two successive
vectors $\vec{v}_i$ and $\vec{v}_{i+1}$, where 
$\vec{v}_i=\vec{r}_{i+1}-\vec{r}_i$. The dihedral angle, $\phi_i$, is the
angle between two vector products $\vec{v}_{i-1}\times\vec{v}_i$ and
$\vec{v}_i\times\vec{v}_{i+1}$. 
The angular distances to the native state have
not been studied in previous works.

We define the structural susceptibilities corresponding to
the distances (1) as follows
\begin{eqnarray}
\chi_d \; &=& \; < \delta^2_{d} > - < \delta_{d} >^2 \, , \nonumber\\
\chi_{ba} \; &=& \; < \delta^2_{ba} > - < \delta_{ba} >^2 \, , \nonumber\\
\chi_{da} \; &=& \; < \delta^2_{da} > - < \delta_{da} >^2 \; ,
\end{eqnarray}
where the angular brackets indicate a thermodynamic average. 
As one can see below, these three susceptibilities behave
qualitatively in the same way. The sharpness of their peaks may be,
however, different (see, for instance, Fig. 4) and it is useful
to calculate all of them.

In the case of an off-lattice model the departures of the sequence
geometry from its native conformation are usually described through the
structural overlap function \cite{Camacho} as
\begin{equation}
\delta_o \; = \; 1 - \frac{2}{N^2-3N+2} \sum_{i \neq j,j\pm 1}
\Theta(\delta_c-|d_{ij}-d_{ij}^{NAT}|)) \; ,
\end{equation}
where $\Theta(x)$ is the Heaviside function.
The overlap structural susceptibility,
$\chi_o$, is then defined as the
thermal fluctuation of $\chi_s$   
\begin{equation}
\chi_o \; = \; < \delta^2_{o} > - < \delta_o >^2 \; .
\end{equation}
The maximum in $\chi_o$, when plotted against $T$, may be interpreted as
a signature of the folding temperature $T_f$ \cite{Camacho,Thirumalai}.
The advantage of the new definitions of the structural susceptibilities (2)
compared to $\chi_o$ is that the native basin $\delta_c$ is not involved in
their computation. 

We have also studied the following derivatives of distances with respect
to $T$
\begin{eqnarray}
D_d \; = \; \frac{d<\delta_d>}{dT} \; ,
D_{ba} \; = \; \frac{d<\delta_{ba}>}{dT} \; ,\nonumber\\
D_{da} \; = \; \frac{d<\delta_{da}>}{dT} \; ,
D_o \; = \; \frac{d<\delta_o>}{dT} \; ,\nonumber\\
D_g \; = \; \frac{d<R_g>}{dT} \; ,
\end{eqnarray}
where $R_g$ is the gyration radius. 
Naively one can expect that the peaks of the derivatives $D$, when plotted
against $T$, would coincide with those of the corresponding susceptibilities
$\chi$.
It is, however, true only when the native conformations are compact.

Using the Monte Carlo and the
molecular dynamics simulations we have demonstrated
that $T_f$ locates at  the peaks
of $\chi_d$ ($D_d$), $\chi_{ba}$ ($D_{ba}$) or $\chi_{da}$ ($D_{da}$)
provided the native conformations
are compact. 
Thus, the determination of
$T_f$ does not require the native basin $\delta_c$.
This is the main advantage of the new quantities given by Eqs. (2) and (5).

The situation becomes more complicated when
the native conformations are not compact.
In this case the native basin is necessary 
for the accurate estimate of $T_f$. The information about
the foldability may be, however, obtained without $\delta_c$ 
monitoring the temperature dependence of $D_d, D_{ba}$ and $D_{da}$.
Namely, a good folder would have only one peak in $D_d, D_{ba}$ or $D_{da}$,
when plotted against temperature, whereas a bad folder would have two peaks.
This may serve as a criterion to distinguish the good folders from the bad
ones.

We focus on four sequences whose native conformations are 
shown in Fig.1. The 27-monomer lattice chain, 
$L_{27}$, is a Go sequence \cite{Go}. Its Hamiltonian
is as follows
\begin{equation}
H \; \; = \; \; \sum_{i<j} \alpha_{ij} \Delta_{ij} \; ,
\end{equation}
where $\Delta_{ij}=1 $ if monomers $i$ and $j$ are in contact and
$\Delta_{ij}=0$ otherwise. The quantity $\alpha_{ij}=-1$ if monomers $i$ 
and $j$ are in contact in the native conformation and 
$\alpha_{ij}=0$ otherwise.
We use $L_{27}$ to check the behavior of the new quantities 
$\delta_{ba}$ ($D_{ba}$)
and $\delta_{ha}$ ($D_{ha}$) for the lattice models.

The sequences denoted by $G$ and $R'$ are
two-dimensional
versions of the model introduced by Iori, Marinari and Parisi \cite{IMP}.
The Hamiltonian is given by
\begin{equation}
H\; \; = \; \; \sum_{i \neq j} \{ k (d_{i,j} - d_0')^2 \delta_{i,j+1}
+ 4 \epsilon [ \frac{C}{d_{i,j}^{12}} - \frac{A_{ij}}{d_{i,j}^6} ] \} \; \; ,
\end{equation}
where $i$ and $j$ range from 1 to 
$N$=16. 
$d_{ij}$ is measured in units of $\sigma$, the typical value of which is
$\sigma=5\r{A}$. We take $d_0'$ to be equal to $2^{1/6}\sigma$ and 1.16$\sigma$
for $G$ and $R'$, respectively \cite{Li}.
The harmonic term
in the Hamiltonian, with the spring constant $k$,
couples the beads that are adjacent along the chain.  The remaining terms 
represent the Lennard-Jones potential. 
Random values of $A_{ij}$ describe the quenched disorder. 
In Eq. (7) $\epsilon$ is the typical Lennard-Jones
energy parameter. We adopt the units in which $C$=1 and
consider $k$ to be equal to 25$\epsilon$.
Smaller values of $k$ may violate the self-avoidance of the chain.
The coupling constants $A_{ij}$ for system R' are listed in Ref. \cite{Li}.
These are shifted Gaussian-distributed numbers with the strongest attracting
couplings assigned to the native contacts. For system G, $A_{ij}$ is taken
to be 1  or 0 for the native and non-native contacts respectively.
System R' has been shown to be structurally overconstrained and hard to fold.

\begin{figure}
\epsfxsize=3.4in
\centerline{\epsffile{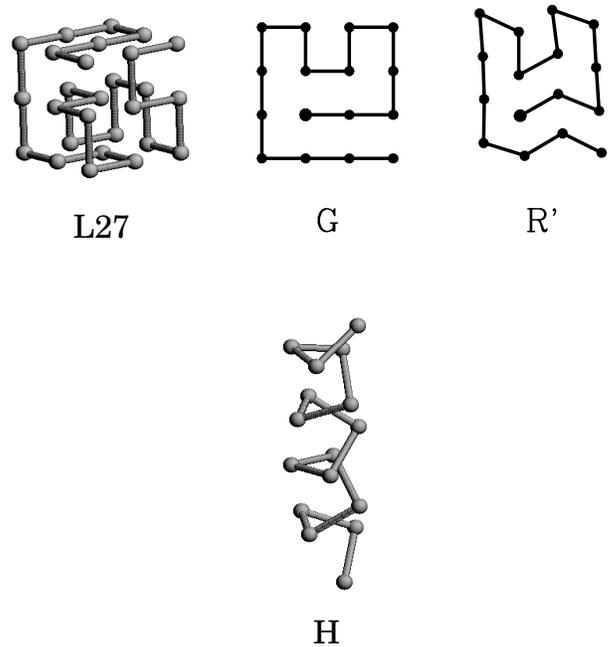}}
\vspace{0.2in}
\caption{Native conformations of four sequences studied
in this work.}
\end{figure}

The helical system $H$ has a native state that mimics 
typical $\alpha$-helix secondary structures.
In this case the distances between beeds are assumed to have the
length $d_0=3.8 \r{A}$. 
As one proceeds along the helix axis from one bead to another,
the bead's azimuthal angle is rotated by 100$^o$ and the azimuthal length
is displaced by 1.5 $\r{A}$. The Hamiltonian used to describe the
helix is a Go-like modification of Eq. (7) and it reads \cite{Hoang}
\begin{equation}
H \; \; = \; \; V^{BB} \; + \; V^{NAT} \; + \; V^{NON} \; .
\end{equation}
The first term is a backbone potential which includes the harmonic and 
anharmonic interactions
\begin{equation}
V^{BB} \; = \; \sum_{i=1}^{N-1} [ k_1 (d_{i,i+1} -d_0)^2 
+  k_2 (d_{i,i+1} -d_0)^4 ] \; .
\end{equation}
We take $d_0=3.8 \r{A}$, $k_1=\epsilon$ and $k_2=100\epsilon$.
The interaction between residues which form native contacts in the
target conformation is chosen to be of the Lennard-Jones form
\begin{equation}
V^{NAT} \; = \; \sum_{i<j}^{NAT} 4 \epsilon [(\frac{\sigma _{ij}}{d_{ij}})^{12}
- (\frac{\sigma _{ij}}{d_{ij}})^{6} ] \, .
\end{equation}
We choose $\sigma _{ij}$ so that each contact in
the native structure is stabilized at the minimum of the potential,
i. e. $\sigma _{ij} = 2^{-1/6} d_{ij}^N$, where $d_{ij}^N$ is the 
length of the corresponding native contact.
Residues that do not 
form the native contacts interact via a repulsive soft core 
potential $V^{NON}$, where
\begin{eqnarray}
V^{NON} \; \; = \; \; \sum_{i<j}^{NON} \; V_{ij}^{NON} \; , \\
V_{ij}^{NON}= \left\{ \begin{array}{r@{\quad \quad}l}
4 \epsilon [(\frac{\sigma _0}{d_{ij}})^{12}-(\frac{\sigma _0}{d_{ij}})^6 ]+
\epsilon &
, d_{ij}<d_{cut}\\
0 & , d_{ij}>d_{cut}.
\end{array} \right. 
\end{eqnarray}
Here $\sigma _0=2^{-1/6} d_{cut}, d_{cut}=5.5 \r{A}$.
The difference between a Go and Go-like sequences is in
the choice of the non-native contact interaction energy
which is taken to be zero
for the Go sequence and nonzero for the latter one.

\begin{figure}
\vspace{0.2in}
\epsfxsize=4in
\centerline{\epsffile{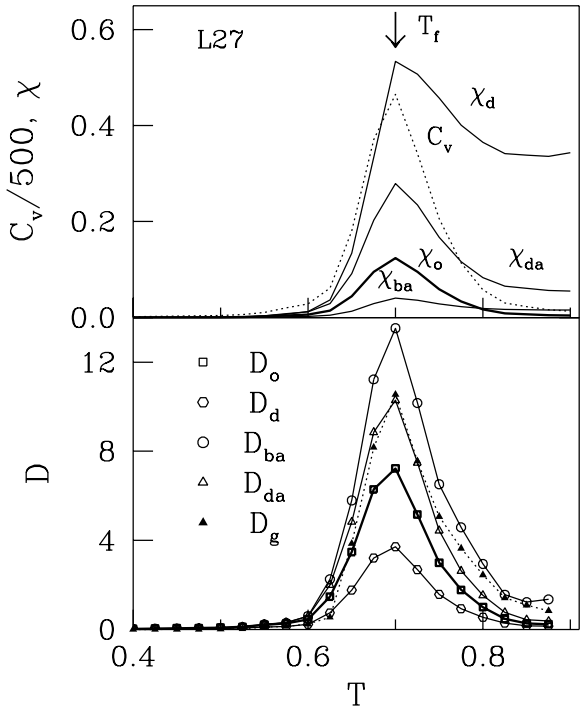}}
\vspace{0.2in}
\caption{The temperature dependence of $C_v$, $\chi$ and $D$ for sequence
L27. The arrow corresponds to $T_f$.
$T_f$ is defined as a temperature
at which the probability of being in the native state is $1/2$.
$C_v$ and $D_g$ are denoted by dotted lines whereas $\chi_o$ and
$D_o$ - by thick lines.
The results are averaged over 50 starting conformations.
The error bars are smaller than the symbol sizes.}
\end{figure}

The thermodynamics of $L_{27}$ is studied by a Monte Carlo procedure
that satisfies the detailed balance condition \cite{Cieplak,Chan}.
The dynamics
allows for single and two-monomer (crankshaft) moves. For each
conformation of the chain one has $A$ possible moves and the maximum
value of $A$, $A_{max}$, is equal to $A_{max}=N+2$. In our 27-monomer case
$A_{max}=29$. For a conformation with $A$ possible moves, the probability
to attempt any move is taken to be $A/A_{max}$ and the probability not to do
any attempt is 1-$A/A_{max}$ \cite{Chan}. In addition, probability
to do a single move is reduced by the factor of 0.2 and to do
the double move - by 0.8 \cite{Chan}.
The attempts are rejected or accepted as in the standard Metropolis
method. The equilibration is checked by monitoring the stability of data
against at least three-times longer runs. We have used typically
$10^6$ Monte Carlo steps
(the first $5\times 10^5$ steps are not taken into account when 
averaging).

\begin{figure}
\vspace{0.2in}
\epsfxsize=4in
\centerline{\epsffile{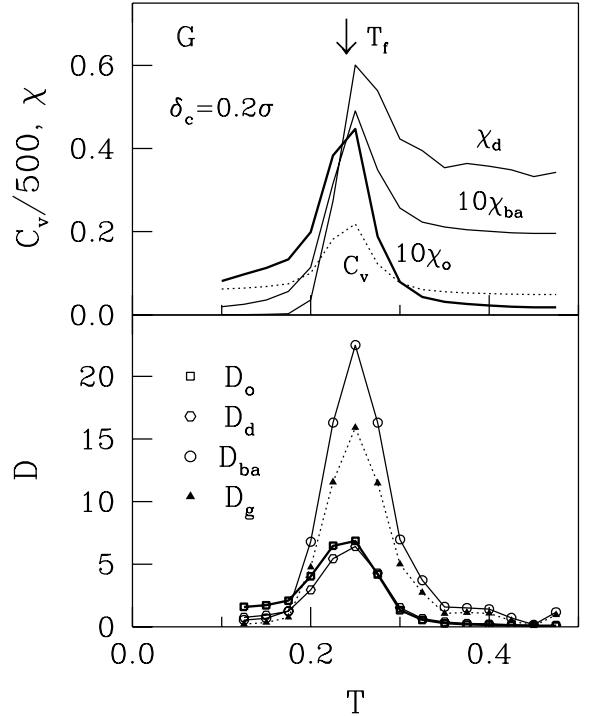}}
\vspace{0.2in}
\caption{The same as in Fig. 2 but for sequence
G. The native basin defined by the shape distortion approach
[7] is equal to $\delta_c=0.2\sigma$.
The results are averaged over 100 molecular dynamics trajectories.}
\end{figure}

Fig. 2 shows the temperature dependence of $C_v$, $\chi$ and $D$
for $L_{27}$, where $\chi$ ($D$) is a common notation for $\chi_d$ ($D_d$),
$\chi_{ba}$ ($D_{ba}$), $\chi_{da}$ ($D_{da}$)
and $\chi_o$ ($D_o$). In this on-lattice case 
the overlap structural susceptibility 
$\chi_o$
is also given by Eq. (4)
but $\delta_o$ reads as follows \cite{Camacho}
\begin{equation}
\delta_o \; = \; 1 - \frac{2}{N^2-3N+2} \sum_{i \neq j,j\pm 1}
\delta(d_{ij}-d_{ij}^{NAT}) \; .
\end{equation}
For sequence $L_{27}$ the peaks of all 
quantities are located at $T=T_f$. The fact that the maxima of $\chi_o$ and
$D_g$  are located at the same position has been also observed 
for some on-lattice sequences \cite{Klimov2}. 
Our new result
is that, similar to $\chi_o$,
the susceptibilities based on the fluctuations of
the distances to the native conformation  and $D$
also give a correct position for $T_f$. 
According to the thermodynamic criterion \cite{Camacho,Klimov1},
$L_{27}$ should be a good  folder because
$T_f$ coincides with the collapse temperature $T_{\theta}$
($T_{\theta}$ is defined as a temperature where
$C_v$ develops a peak).

In order to study
the time evolution  of the off-lattice sequences
$G, R'$ and $H$,
we use the equations of motion by
the Langevin uncorrelated noise terms:
\begin{equation}
m\ddot{{\bf r}} = -\Gamma \dot{{\bf r}} + F_c + \eta \;,
\end{equation}
where $F_c=-\nabla_r E_p$ and
\begin{equation}
\left<\eta(0)\eta(t)\right> = 2\Gamma k_B T \delta(t) .
\label{eqgam}
\end{equation}
Here $k_B$ and $\Gamma$ are the Boltzmann constant and the kinetic coefficient,
respectively.
Equation (14) is integrated
by the fifth
order predictor-corrector method \cite{MD}. The integration step
is chosen to be 0.005$\tau$, where $\tau=m \sigma ^2 / \epsilon$
is the characteristic time unit and $m$ is the mass of a bead.
We take $\Gamma $ equal to 2. In the following, the temperature will be
measured in the reduced units of $\epsilon/k_B$.

\begin{figure}
\vspace{0.2in}
\epsfxsize=4in
\centerline{\epsffile{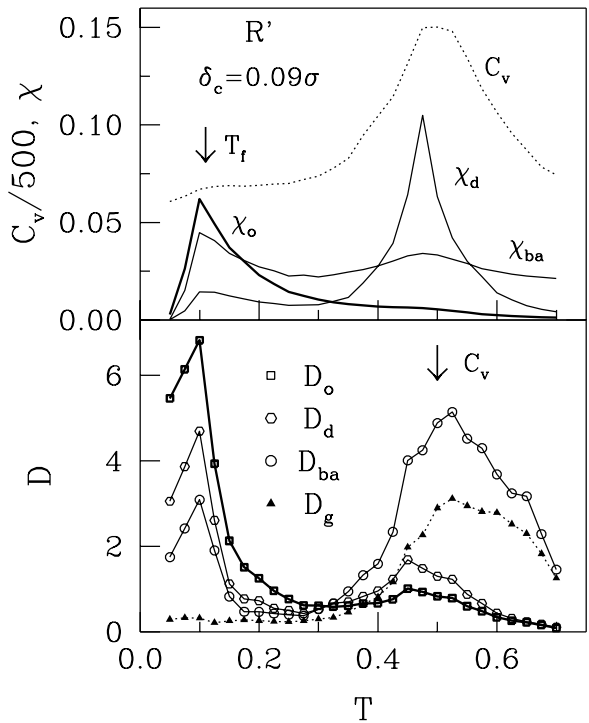}}
\vspace{0.2in}
\caption{The same as in Fig. 2 but for sequence
R'. The native basin is equal to $\delta_c=0.09\sigma$.
The results are averaged over 170 molecular dynamics trajectories.}
\end{figure}

The folding properties of $G$, $R'$ and $H$ were characterized in detail
previously \cite{Li,Hoang}. One of them, $R$', is a bad folder
and two others are good folders.
We calculate the thermodynamic quantities of $G$, $R'$ and $H$
by averaging over many molecular
dynamics trajectorjes using the native state as the starting configuration
to make sure that the evolution takes place in the right part of the phase
space \cite{Li}. For all of these sequences, the time used for averaging in 
each trajectory is 4000$\tau$ for each temperature. The first 2000$\tau$ are
discarded. 

Fig. 3 and 4 show the results for $G$ and $R'$. 
Since these sequences are two-dimensional, $\chi_{ha}$ and $D_{ha}$
corresponding to the dihedral angles do not appear.
The basin was obtained by the shape distortion approach \cite{Li1} and
is equal to $\delta_c=0.2\sigma$ and $\delta_c=0.09\sigma$ for $G$
and $R$', respectively.
Within the error bars of 0.02 all of the 
maxima of $\chi$, $D$ and $C_v$ are located at the
folding temperature $T_f$  ($T_f=0.24 \pm 0.02$ and $T_f=0.10 \pm 0.02$
for $G$ and $R$', respectively).
Therefore, the determination of $T_f$ does not require the native basin
because it is enough
to find the peak of $\chi$ (or of $D$) in which $\delta_c$ is not
involved.

For $R$', $\chi _o$ has only one peak at $T_f$ whereas $\chi_d$ and
$\chi_{ba}$ have an additional maximum at $T=T_{\theta}$. Therefore,
the advantage
of $\chi_d$ and $\chi_{ba}$ compared to $\chi _o$ 
is that they allow to find not only $T_f$ but also  
$T_{\theta}$. Since the maximum of $D_g$
is broad around the folding temperature, it is better 
to locate $T_f$ as a second peak of $\chi_d$ ($D_d$) or $\chi_{ba}$
($D_{ba}$). It demonstrates the another advantage of the new quantities
compared to the standard quantity $D_g$.

It should be noted that the behavior of $\chi_d$ and $\chi_{ba}$ is 
qualitatively the same but there is a quantitative difference 
in the sharpness of their peaks. It is clear from Fig. 4
that at $T=T_f$ the maximum of $\chi_{ba}$ is more pronounced compared
to that of $\chi_d$. An opposite situation takes place at $T=T_{\theta}$.
So, the study of all of susceptibilities would help us to isolate 
peaks better.

The fact that $\chi_0$ has only one peak, but the others 
have two may be explained in the following way. Since $\chi_0$ is a
fluctuation of the overlap with the native state it reflects the    
behavior of the system
in the vicinity of the native basin and it should have, therefore,
only one peak at $T_f$. The remaining susceptibilities related to the
chain compactness would have two maxima at $T_f$ and $T_{\theta}$ where
the topology changes abruptly.

The temperature dependence of $\chi$, $D$ and $C_v$ for the
three-dimensional sequence $H$ is shown in Fig. 5. 
In this case we have the basin $\delta_c=0.12\sigma$ and $T_f=0.24 \pm 0.02$
\cite{Hoang,Li2}.
Since $\chi_d, \chi_{ba}$ and $\chi_{da}$ do not display any peak
in the relevant temperature interval, they cannot be used to
determine $T_f$. It is also true for $D$ (their extremal points
are located at $T=0.325 \pm 0.025$
which is far from $T_f$). 
The overlap susceptibility  $\chi_o$ has the maximum 
at $T_{\chi_o}=0.275 \pm 0.025$. Within the error bars $T_{\chi_o}$ may be 
identified as $T_f$ but such an estimate is less accurate compared
to the case of $G$ and $R$'.
Furthermore its computation involves the native basin $\delta_c$.

From the results presented in Figs. 2 - 5 we 
propose the following criterion for foldability:
a good folder would have only one peak
in the derivatives of distances to the native state with respect to
temperature whereas  a bad folder has two. 
Our criterion is compatible with the fact that for the good folders the folding 
takes place just after the collapse transition. A three state
scenario of folding is, however,
more suitable for the bad folders \cite{Thirumalai}.
Thus, one can still gain information about the foldability
for $H$ without the native basin $\delta_c$.

The question we ask now is why $H$ is so different from the other
sequences. The main difference
is that its native conformation is not compact.
It results in the non-trivial dependence of $R_g$ on $T$: $D_g$
does not develop a maximum but rather a minimum around the the collapse
transition. This leads to the anomal behavior of
$\chi_d, \chi_{ba}$ and $\chi_{ha}$.

\begin{figure}
\vspace{0.2in}
\epsfxsize=4in
\centerline{\epsffile{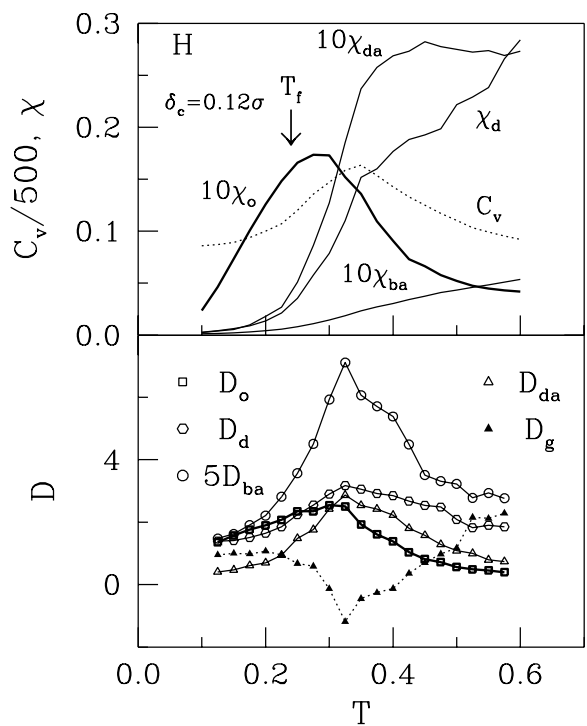}}
\vspace{0.2in}
\caption{The same as in Fig. 2 but for sequence
H. The native basin is equal to $\delta_c=0.12\sigma$.
The results are averaged over 200 molecular dynamics trajectories.}
\end{figure}

In conclusion, we have introduced 
several new structural
susceptibilities as fluctuations of distances to
the native conformation. If the native confomation of proteins
is compact $T_f$
may be obtained from the peak of $\chi$
and the native basin is not required. For sequences with non-compact
native conformation $\delta_c$ is not needed to establish the
foldability but the accurate estimate of $T_f$ should involve it.
The number of peaks in the derivatives of distances to the
native state with respect to temperature, when plotted against $T$, may serve as a tool to distinguish 
between good and bad folders.
The question of why the susceptibilities $\chi$ and the derivatives $D$
behave in the same way if the native conformations are compact remains
to be elucidated. Nevertheless, in agreement with other studies 
(see, for instance, Ref. \cite{Baker} and references there),
our results indicate that the topology
of the native state plays a crucial role in the folding process.

We thank M. Cieplak for useful discussions.
This work was supported by Komitet Badan Naukowych (Poland; Grant number\
2P03B-146 18).

\vspace{1cm}


\end{document}